\documentclass[a4paper]{jpconf}
\usepackage{graphicx}
\usepackage{natbib}
\usepackage{citesort}

\begin{document}

%\title{Exceptional brightening of J1430+4204 at z=4.72 }
\title{The compact radio structure of the high-redshift blazar J1430+4204 before and after a major outburst}
\author{P\'eter Veres$^{1,2}$, S\'andor Frey$^{3,5}$, Zsolt Paragi$^{4,5}$, Leonid I. Gurvits$^{4}$}

\address{$^1$Dept. of Physics of Complex Systems, E\"otv\"os University, P\'azm\'any P. s. 1/A, H-1117 Budapest, Hungary}
\address{$^2$Dept. of Physics, Bolyai Military University, POB 15, Budapest, H-1581, Hungary}
\address{$^3$F\"OMI Satellite Geodetic Observatory, POB 585, H-1592 Budapest, Hungary}
\address{$^4$Joint Institute for VLBI in Europe, POB 2, 7990 AA Dwingeloo, The Netherlands}
\address{$^5$MTA Research Group for Physical Geodesy and Geodynamics, POB 91, H-1521 Budapest, Hungary}
\ead{veresp@elte.hu}

\begin{abstract}
The high-redshift ($z=4.72$) blazar J1430+4204 produced an exceptional radio outburst
in 2006. We analyzed 15-GHz radio interferometric images obtained with the Very Long Baseline Array (VLBA) 
before and after the outburst, to search for possible structural changes on milli-arcsecond angular scales 
and to determine physical parameters of the source.
\end{abstract}

\section{Introduction}
Active galactic nuclei (AGNs) are thought to harbor supermassive (up to
$\sim10^{10}M_{\odot}$) black holes. Accretion onto these black holes is
responsible for the extreme luminosity of AGNs over the whole electromagnetic
spectrum. Part of the infalling matter may be transformed into jets ejected
with relativistic speeds. The radio emission in radio-loud AGNs originates from
these jets via synchrotron process. If a jet points close to the line of sight
toward the observer, its brightness is significantly enhanced. For a review of
the unified model of AGNs see Urry \& Padovani (1995).
%\cite{1995PASP..107..803U}.

A particular class of AGNs are blazars. They show large variations in
brightness from the radio to the gamma-ray regime. They have no emission lines
characteristic to other AGNs in the optical spectrum. According to the physical
models of AGN (Urry \& Padovani 1995), we see at these objects almost exactly
in the direction of the jet. 
%\citep{1995PASP..107..803U}

J1430+4204 (B1428+4217) is a blazar with a flat radio spectrum at an extremely
high redshift, $z=4.72$ (Hook \& McMahon 1998). Radio flux density
monitoring at 15~GHz revealed a significant brightening of J1430+4204, starting
in 2004 and reaching its peak in 2006 (Worsley et al. 2006).  The object
increased its flux density by a factor of 3 in about 4 months (in the source
rest frame).
%\cite{1998MNRAS.294L...7H}
%\citep{2006MNRAS.368..844W}

The radio structure of J1430+4204 at the milli-arcsecond (mas) scale is predominantly
compact as revealed by Very Long Baseline Interferometry (VLBI) imaging
observations (e.g. Paragi et al. 1994, Helmboldt et al. 2007).  A weak
extension to the bright compact core is also seen in the W-SW direction.
%\cite[e.g.][]{1999A&A...344...51P,2007ApJ...658..203H}

Total flux density outbursts are usually followed by an emergence of a new jet
component in the VLBI images of radio AGNs.  By observing J1430+4204 after the
brightening, we aimed at detecting a new component in a hope to have a
zero-epoch point for a later measurement of its apparent proper motion. A study
of jet kinematics at such a high redshift would have been interesting since the
best-observed sample in the 15-GHz MOJAVE survey (Lister et al. 2007)
%\cite{2009AJ....138.1874L} 
is restricted to $z<3.5$.   

\section{Observations and data processing}
\begin{figure}[h]
\centering
\includegraphics[width=54mm,angle=270,bb=50 80 520 770, clip= ]{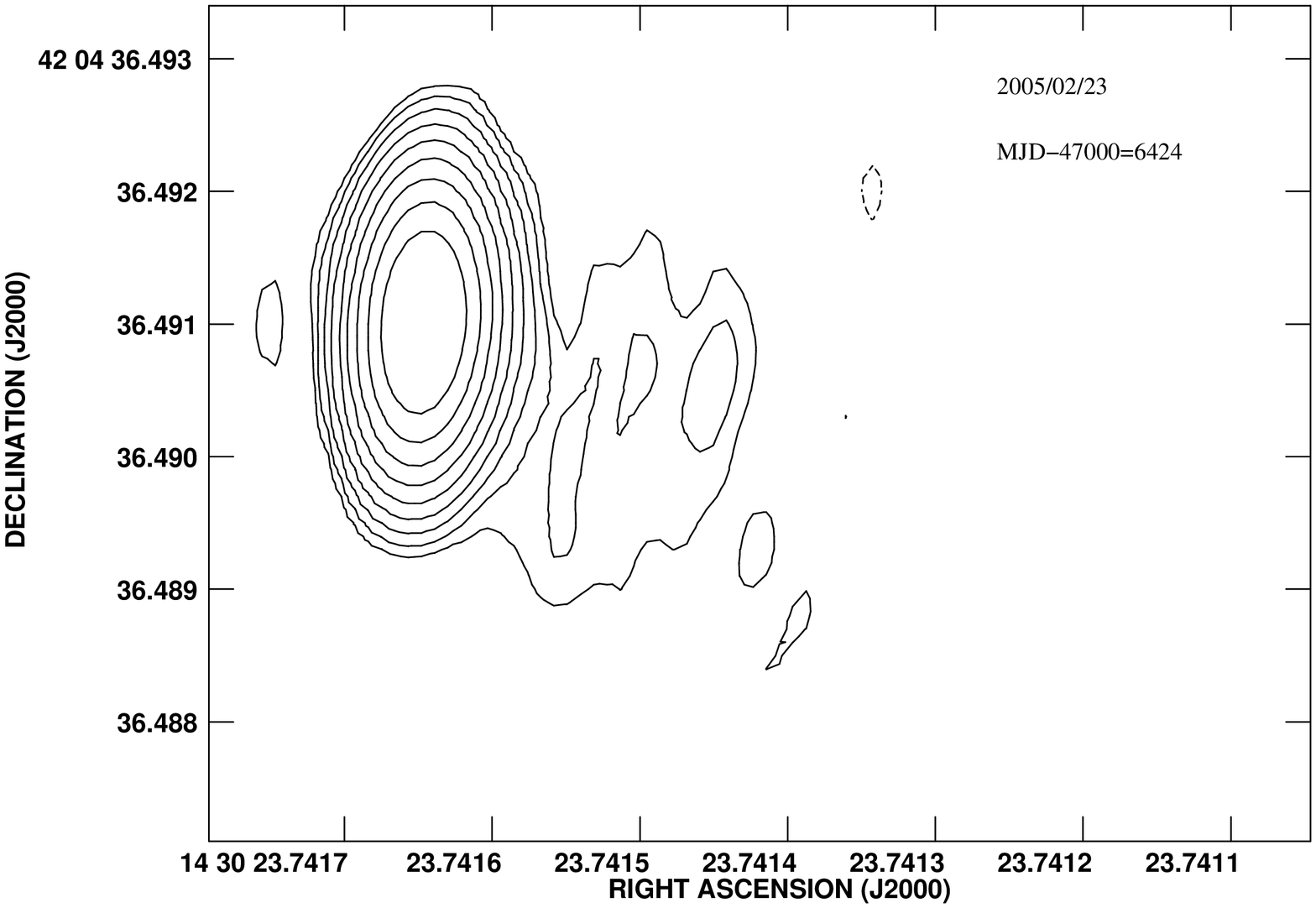}
\includegraphics[width=54mm,angle=270,bb=50 80 520 770, clip= ]{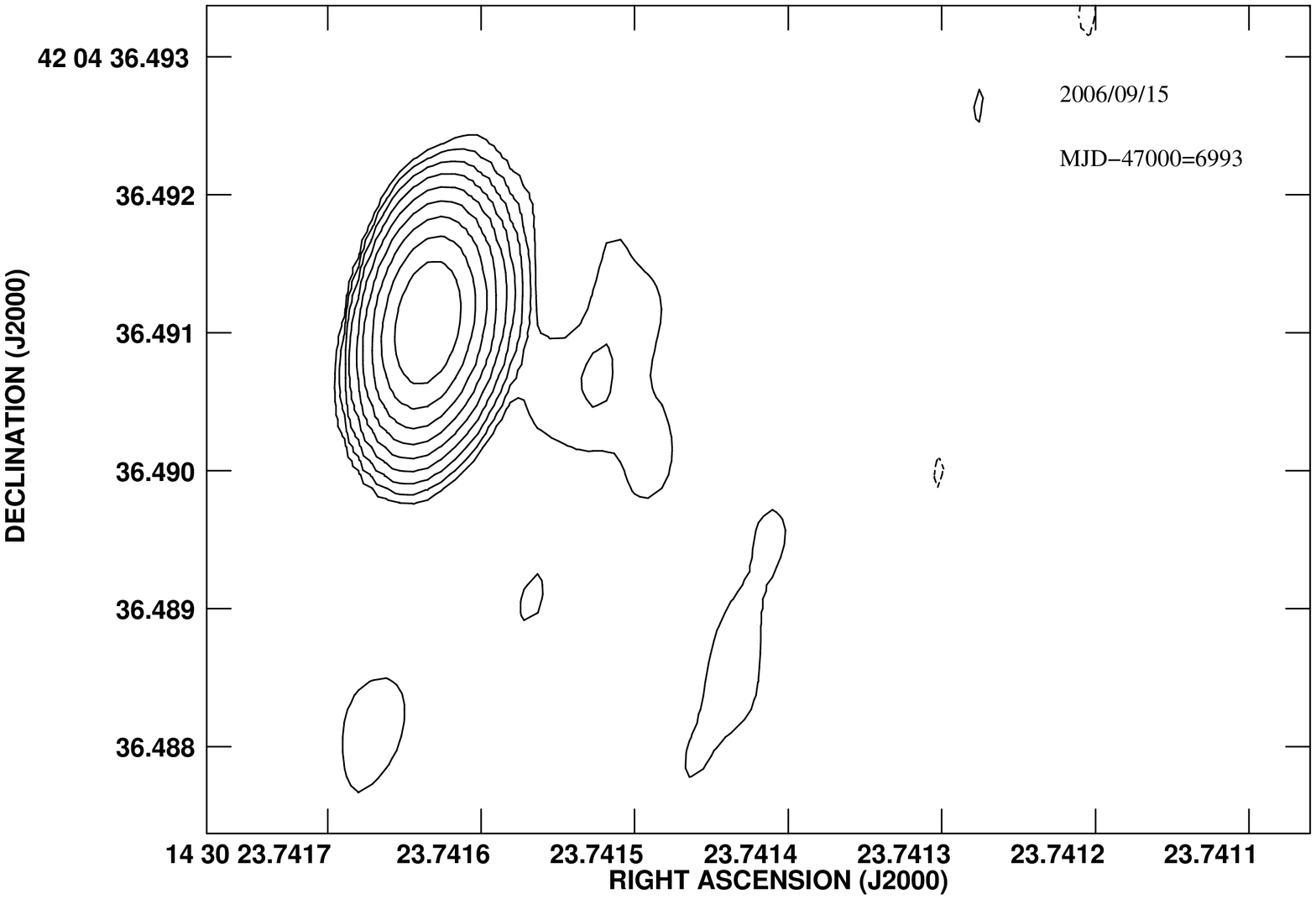}

\caption{15-GHz VLBA images of J1430+4204 on 23 Feb 2005 {\it (left)} and 15
Sep 2006 {\it (right)}. In the left image, the peak brightness is 200 mJy/beam,
the restoring beam is 1.17~mas$\times$0.52mas at the position angle
PA=-3.8$^{\circ}$. In the right image, the peak brightness is 159 mJy/beam,
the restoring beam is 1.22~mas$\times$0.58 mas at PA=-11.95$^{\circ}$. In both
cases, the lowest contour levels are at 0.3 mJy/beam, the positive contours
increase by a factor of 2.}

\label{images}
\end{figure}

We observed J1430+4204 for 8 hours at 15~GHz with the ten 25-m diameter radio
telescopes of the NRAO\footnote{The National Radio Astronomy Observatory is a
facility of the National Science Foundation operated under cooperative
agreement by Associated Universities, Inc.} Very Long Baseline Array (VLBA) on
15 Sep 2006.  We also found a similar full-polarization experiment (code BY019)
in the NRAO data archive.  These observations were done on 23 Feb 2005, prior
to the brightness peak.  We performed standard VLBI calibration, imaging and
model-fitting procedures for both data sets using the NRAO Astronomical Image
Processing System (AIPS) and the Caltech Difmap program.  Our total intensity
images are displayed in Fig.~\ref{images}.  The fractional linear polarization
of the VLBI core was $\sim1\%$ and $2\%$ at the first and second epoch,
respectively.

\begin{figure}[h]
\centering
\includegraphics[width=35pc]{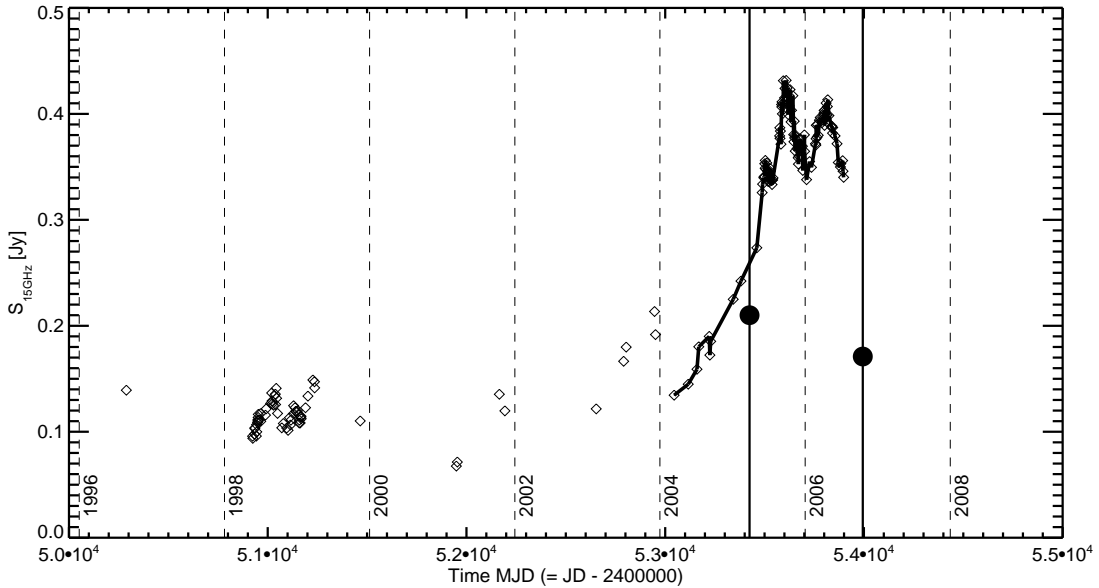}

\caption{The 15-GHz flux density vs. time for J1430+4204 from the Ryle Telescope
monitoring (G. Pooley, priv. comm.). The points are connected in the flaring
phase for clarity. Solid vertical lines mark the times of the two VLBA
observations, the filled circles correspond to the measured VLBI flux
densities. Dashed lines mark calendar years as indicated.}

\label{lightcurve} 
\end{figure}

The source total flux density at 15~GHz was monitored at the Ryle Telescope (Fabian et al. 1999). 
The light curve is shown in Fig.~\ref{lightcurve}. 
%The times of the two VLBA observations, 
%as well as the sum of the flux densities in the VLBI components are marked. 
%\cite{1999MNRAS.308L...6F}

%---------------- VLBA images

%---------------- Ryle Tel. light curve

\section{Determination of the source parameters}

We used three methods for computing the following parameters of J1430+4204: the
apparent speed of a possible blob in the jet, $\beta_{app}= \frac{\beta \sin
\theta}{1-\beta \cos \theta}$; the Doppler factor,   $\delta  =
\frac{1}{\Gamma(1-\beta \cos \theta)}$; and the Lorentz factor,
$\Gamma=\frac{\beta_{app}^2 + \delta^2+1}{2 \delta^2} = (1-\beta^2)^{-1/2}$.
Here $\beta<1$ is the bulk speed of the material in the jet, expressed in the
units of the speed of light $c$.  For the jet angle to the line of sight,
$\theta=\arctan \left(\frac{2 \beta_{app}}{\beta_{app}^2 + \delta^2-1}\right)
$, we assumed $3^{\circ}$ as found from bulk Comptonization modeling of the
observed X-ray spectrum of J1430+4204 (Cellotti et al. 2007)
%\cite{2007MNRAS.375..417C}
.

The most important physical parameter characterizing the jet flow is the bulk
Lorentz factor ($\Gamma$).  We are also interested in the apparent tangential
velocity of the putative blob in the jet which we translate into proper motion.
The distance scale at this redshift is $6.588$ pc/mas (assuming a cosmological
model with $H_{\rm 0} = 0.71$~km~s$^{-1}$~Mpc$^{-1}$, $\Omega_{\Lambda}=0.73$
and $\Omega_{\rm M}=0.27 $).

\subsection{Parameters from the radio variability}
We fitted an exponential function to the brightest flare in the flux density
curve (Fig.~\ref{lightcurve}) and determined the variability brightness
temperature (Hovatta et al. 2009). Furthermore, we took the intrinsic
brightness temperature as $T_{\rm b,int}=5\times 10^{10}$~K, which assumes
equipartition between the particles and the magnetic field in the
radio-emitting region (Readhead 1994).  We then calculated the so-called
variability Lorentz factor and the apparent speed. 

%\cite{2009A&A...494..527H} \citep{1994ApJ...426...51R}

%\begin{equation}
%\Delta S(t)= \left\{
%\begin{array}{ll}
%\Delta S_{max} e^{(t-t_{max})/\tau},\quad t<t_{max} \\
%\Delta S_{max} e^{(t-t_{max})/-1.3\tau},\quad  t>t_{max}
%\end{array}\right.
%\end{equation}
%\begin{eqnarray*}
%and $ T_{b,var} = 1.548\times 10^{-32}\frac{\Delta S_{max} d_L^2}{\nu^2 \tau^2 (1+z)}$
%\delta_{var} &=& \left[{\frac{T_{b,var}}{T_{b,int}}}\right]^{1/3}\\
%T_{b,int} &\simeq& 5 \times 10^{10} K 
%\end{eqnarray*}

%\begin{figure}[h]
%\begin{center}
%\includegraphics[width=28pc]{fit2csucsPUB.ps}\hspace{2pc}%
%\begin{minipage}[b]{36pc}\caption{\label{fit} The rising phase of the flare fitted with a two-component model.}
%\end{minipage}
%\end{center}
%\end{figure}

We obtained $\Delta S = 412$~mJy for the flare amplitude and $\tau = 458.4^{\rm
d}$ for the rise time.  These led to the variability brightness temperature
$T_{\rm b,var} = 4.1\times 10^{13}$~K, $\delta_{\rm var} = 9.4$, $\beta_{\rm
app,var} = 2.4$, $\Gamma_{\rm var} = 5.1$. The apparent proper motion of the
jet component is $\mu = 0.03$~mas/year.  This means that during the time
between the two VLBA observations ($\Delta T \simeq 570^{\rm d}$) the supposed
blob had moved by $\simeq 0.05$ mas. This angular displacement is below the
limit which we can possibly detect with the VLBA at this wavelength ($\sim 0.1$
mas).

\subsection{Parameters from the brightness temperature measured with VLBI}
We fitted circular Gaussian brightness distribution models to the VLBI visibility 
data at $15$~GHz at both epochs. This gave us $T_{\rm b}\simeq1.2\times 10^{12}$~K brightness temperatures.
It corresponds to a Lorentz factor of $\Gamma \simeq 10$, if we again assume the equipartition value 
for the intrinsic brightness temperature $T_{\rm b,int}$. This method
also predicts a proper motion of $\mu = 0.03$ mas/year. 
%\begin{eqnarray*} 
%The brightness temperature will be $T_{b,obs} = 1.22\times 10^{12}
%\frac{S}{\theta_{\mathrm{Gauss}}^2 \nu^2} (1+z)$ and the Doppler factor: $\delta_{EQ} =
%\frac{T_{b,obs}}{T_{b,int}} $
%\end{eqnarray*}

\subsection{Parameters from the inverse Compton process}
By assuming that the observed X-ray flux of an AGN is of the inverse Compton
origin, we can estimate the Doppler factor (Guijosa \& Daly 1994) . For
J1430+4204, we took the X-ray data from Celotti et al. (2007) . The formula for
the Doppler factor is 
%\cite{1996ApJ...461..600G} \cite{2007MNRAS.375..417C}
\begin{equation}
\delta_{\rm IC} = f(\alpha) (1+z) S_{\rm m} \times\left[\frac{\ln(\nu_{\rm b}/\nu_{\rm op}) \nu_{\rm x}^{-\alpha}}{S_{\rm X} \theta_{\rm d}^{6+4\alpha}\nu_{\rm op}^{5+3\alpha}}\right]^{1/(4+2\alpha)} 
\end{equation}
where $f(\alpha) = -0.08 \alpha + 0.14 $, $S_{\rm m}$ is the radio flux
density, $\nu_{\rm b}$ is the synchrotron high-frequency cutoff assumed to be
$10^5$~GHz. The observed frequency of the radio peak is $\nu_{\rm op}=15$~GHz,
the X-ray flux density is $S_{\rm X}=3.7\times 10^{-7}$~Jy taken at $\nu_{\rm
X}=5.2$~keV, $\alpha=-0.4$ is the optically thin spectral index where the
$S_\nu \propto \nu^{\alpha}$ convention is used, and $\theta_{\rm d}$ is the
angular diameter of the source in mas. This third method also provided $\mu
\simeq 0.03$ mas/year for the jet component proper motion.

%\cite{1987slrs.work..280M} Marscher

\section{Discussion and conclusion}
The high-redshift blazar J1430+4204 produced an exceptional radio flux density
outburst in 2006 (Fig.~\ref{lightcurve}).  We imaged the source with the VLBA
at 15~GHz after the time of the flux density peak, and also analyzed the
archive VLBA data taken during the rise of the total flux density curve. At
both epochs, the mas-scale radio structure of the source was similar: a compact
core and a weak extension to S-SW (Fig.~\ref{images}). The core could be fitted
with circular Gaussian components with sizes of 0.076~mas and 0.060~mas (full
width at half maximum) on 23 Feb 2005 and 15 Sep 2006, respectively. The
comparison of the total and VLBI flux densities at the first epoch
(Fig.~\ref{lightcurve}) suggests that $\sim50$~mJy could be attributed to the
radio emission extending to more than a few mas. 

Based on our VLBA imaging, we do not detect any new separate jet component to
be associated with the outburst.  Assuming a small jet angle to the line of
sight, we used three different methods to calculate the expected proper motion
of such a component. These gave consistently small values of the proper motion.
We conclude that our time base and angular resolution were insufficient to
distinguish any new blob in the jet.  Our estimates for the bulk Lorentz factor
($\Gamma \simeq 5-10$) are comparable with the typical values found for other
blazars (e.g. Hovatta et al. 2009).
%\cite[e.g.][]{2009A&A...494..527H}.  

\ack{This work was supported by OTKA grants K077795 and K72515.
We are grateful to Guy Pooley for sharing the Ryle Telescope data with us.}

\section*{References}

Celotti A, Ghisellini G and Fabian A C 2007 {\it MNRAS} {\bf 375} 417\\
Fabian A C, Celotti A, Pooley G, et al.  1999 {\it MNRAS} {\bf 308} L6\\
Guijosa A and Daly R A 1996 {\it ApJ} {\bf 461} 600\\
Helmboldt J F, Taylor G B, Tremblay S, et al. 2007 {\it ApJ} {\bf 658} 203\\
Hook I M and McMahon R G 1998 {\it MNRAS} {\bf 294} L7\\
Hovatta T, Valtaoja E, Tornikoski M and Lahteenmaki A 2009 {\it A\&A} {\bf 494} 527\\
Lister M L, Cohen M H, Homan D C, et al.  2009 {\it AJ} {\bf 138} 1874\\
Paragi Z, Frey S, Gurvits L I, et al. 1999 {\it A\&A} {\bf 344} 51\\
Readhead A C S 1994 {\it ApJ} {\bf 426} 51\\
Urry C M and Padovani P 1995 {\it PASP} {\bf 107} 803\\

%\section*{References} \begin{thereferences}
%\begin{thereferences}
%\bibliography{j1430}
%\end{thereferences}
%\end{thereferences}
\end{document}